# Prevalence and citation advantage of gold open access in the subject areas of the Scopus database

Pablo Dorta-González [a,*], Yolanda Santana-Jiménez [b]

[a] Universidad de Las Palmas de Gran Canaria, TiDES Research Institute, Campus de Tafira, 35017 Las Palmas de Gran Canaria, Spain. *E-mail*: pablo.dorta@ulpgc.es

[b] Universidad de Las Palmas de Gran Canaria, Departamento de Métodos Cuantitativos en Economía y Gestión, Campus de Tafira, 35017 Las Palmas de Gran Canaria, Spain. *E-mail*: yolanda.santana@ulpgc.es

[*] Corresponding author and proofs. E-mail: pablo.dorta@ulpgc.es (P. Dorta-González)

**Abstract**

The potential benefit of open access (OA) in relation to citation impact has been discussed in the literature in depth. The methodology used to test the OA citation advantage includes comparing OA vs. non-OA journal impact factors and citations of OA versus non-OA articles published in the same non-OA journals. However, one problem with many studies is that they are small and restricted to one discipline or set of journals-. Moreover, conclusions are not entirely consistent among research areas and 'early view' and 'selection bias' have been suggested as possible explications. In the present paper, an analysis of gold OA from across all areas of research -the 27 subject areas of the Scopus database- is realized. As a novel contribution, this paper takes a journal-level approach to assessing the OA citation advantage, whereas many others take a paper-level approach. Data were obtained from Scimago Lab, sorted using Scopus database, and tagged as OA/non-OA using the DOAJ list. Jointly with the OA citation advantage, the OA prevalence as well as the differences between access types (OA vs. non-OA) in production and referencing are tested. A total of 3,737 OA journals (16.8%) and 18,485 non-OA journals (83.2%) published in 2015 are considered. As the main conclusion, there is no generalizable gold OA citation advantage at journal level.

**Introduction**

The communication of results has benefited greatly from the emergence of the Internet and more scientists are making their research openly accessible to increase its visibility, usage, and citation impact (Björk, 2004; González-Betancor & Dorta-González, 2017). In this respect, open access (OA) was defined in 2002 by *Budapest Open Access Initiative* as free and unrestricted access on the public Internet to literature that scholars provide without expectation of direct payment (Prosser, 2003).

There are two OA modalities: *gold OA* refers to articles in fully accessible OA journals; and *green OA* refers to publishing in a traditional journal, in addition to self-archiving the pre- or post-print paper in a repository (Harnad et al., 2004). With respect to gold OA, the *Directory of Open Access Journals* (DOAJ) is currently the largest index presenting quality controls. According to the DOAJ list, in March 2016 there were 9,389 OA journals, of which 4,989 were totally free and 2,205 had article processing charges (APC). There was no available information about the possible existence of APC for the other 2,195 journals in the DOAJ list.

Many researchers, starting with Lawrence (2001), have found that OA articles tend to have more citations than pay-for-access articles. This OA citation advantage has been observed in a variety of academic fields including computer science (Lawrence, 2001), philosophy, political science, electrical & electronic engineering, and mathematics (Antelman, 2004), physics (Harnad et al., 2004), biology and chemistry (Eysenbach, 2006), as well as civil engineering (Koler-Povh, Južnič & Turk, 2014).

However, since Lawrence proposed in 2001 the OA citation advantage, this postulate has been discussed in the literature in depth without an agreement being reached (Davis et al., 2008; Norris, Oppenheim & Rowland, 2008; Joint, 2009; Gargouri et al., 2010; Wang et al., 2015; Dorta-González et al., 2017). Furthermore, some authors are critical about the causal link between OA and higher citations, stating that the benefits of OA are uncertain and vary among different fields (Craig et al., 2007; Davis & Walters, 2011).

Kurtz et al. (2005), and later other authors (Craig et al., 2007; Moed, 2007; Davis et al., 2008), set out three postulates supporting the existence of a correlation between OA and increased citations: *(i)* OA articles are easier to obtain; and therefore easier to read and cite (*Open Access postulate*); *(ii)* OA articles tend to be available online prior to their publication and therefore begin accumulating citations earlier than pay-for-access articles (*Early View postulate*); and *(iii)* more prominent authors are more likely to provide OA to their articles, and authors are more likely to provide OA to their highest quality articles (*Selection Bias postulate*). Moreover, these authors conclude that early view and selection bias effects are the main factors behind this correlation.

Gaule & Maystre (2011) and Niyazov et al. (2016) found evidence of selection bias in OA, but still estimated a statistically significant citation advantage even after controlling for that bias. Regardless of the validity or generality of their conclusions, these studies establish that any analysis must take into account the effect of time (citation time window) and selection bias.

At journal level, Gumpenberger, Ovalle-Perandones & Gorraiz (2013) showed that the impact factor of gold OA journals was increasing, and that one-third of newly launched OA journals were indexed in the Journal Citation Reports (JCR) after three years. However, Björk and Solomon (2012) argued that the distribution model is not related to journal impact. This result has been confirmed by Solomon, Laakso & Björk (2013), concluding that articles are cited at a similar rate regardless of the distribution model.

One limitation of the existing literature on the subject, whether supporting or refuting the OA citation advantage, is often the small number of articles analysed, the study of a specific scientific area -when important differences in the publication and citation habits existing between areas are well known (Dorta-González et al., 2015; Dorta-González & Dorta-González, 2013)-, and the short citation window considered.

In the present paper, we analyse all the journals for all 27 subject areas of the Scopus database. In order to reduce the early view effect, we consider different citation time windows from two to three years after publication.

We take a journal-level approach to assessing the OA citation advantage, whereas many others take a paper-level approach. This is the novel contribution of the paper.

We address the following questions: *(i)* are there differences between subject areas in the prevalence of OA journals?; *(ii)* are there significant differences between access types in the amount of published and referenced documents in the journals?; and *(iii)* are there significant differences of journal impact between access types?

**Methodology**

In this study, we analyse exclusively gold OA, that is, journals in which all the articles that are published are OA. In this sense, those journals that use a hybrid business model offering the possibility of putting articles in OA when the authors pay the APCs are considered non-OA journals.

The DOAJ list was used as the source for OA tagging. Data were obtained from Scimago Lab, sorted using Scopus database, and tagged as OA/non-OA using the DOAJ list. Therefore, the characteristics of each of these sources might influence the results.

From the 22,256 journals indexed in Scopus in the year 2015, and after excluding some trade journals, the final number of research journals included in the data set was 22,222.

There were about 9,300 journals indexed in the DOAJ list in 2015. From them, a total of 3,737 OA journals were included in the data set, which were all the journals at the same time in DOAJ and Scopus.

The share of journals covered in the study in each thematic area, and in OA/non-OA is shown in Table 1. These proportions are similar to the shares using just Scopus, and therefore the construction of the data set did not distort the underlying situation.

Summarizing, the total number of journals covered in the data set was 22,222; of which 3,737 were OA journals (16.8%) and 18,485 were non-OA journals (83.2%).

Firstly, we performed a descriptive analysis of the OA journal prevalence. Then, we analysed some indicators comparing both OA and non-OA journals, putting especial emphasis on the possible existence of an OA citation advantage. In order to reduce the early view effect, we considered different citation time windows from two to three years after publication.

We considered the following journal indicators. We used "references" to refer to outbound citations from these documents, and "cites" to refer to inbound citations to these documents.

- *Total Documents:* the size of the journal in the current year (2015). All types of documents were considered.
- *References per Document*: average number of bibliographical references per document in the current year (2015).
- *Cites per Document (3 years)*: average citations per document in a 3 year period. This was computed considering the number of citations received by a journal in the current year (2015) to the documents published in the three previous years (2012, 2013, and 2014).
- *Cites per Document (2 years)*: average citations per document in a 2 year period. This was computed considering the number of citations received by a journal in the current year (2015) to the documents published in the two previous years (2013 and 2014). This indicator is comparable to the Journal Impact Factor (JIF), but using all types of documents and the Scopus database instead of Web of Science.
- *H Index*: number of articles (H) in the journal that have received at least H citations. This quantifies both journal production and citation impact. This indicator is not normalized and depends strongly on both the size and the age of the journal.
- *SJR (Scimago Journal Rank)*: average number of weighted citations received in the current year (2015) by the documents published in the journal in the three previous years (2012, 2013, and 2014). This is a measure of scientific influence of journals that accounts for both the number of citations received by a journal and the importance or prestige of the journals where such citations come from (a variant of the eigenvector centrality measure used in network theory).
- *SJR Best Quartile*: best quartile of the journal in the current year (2015) among all subject categories. Scopus uses a journal classification system where each of them is assigned to one or several subject categories. According to the SJR indicator, each journal is placed in a quartile within each category. The Best Quartile is the highest of them all.

In Annexes 1 to 3, the means, medians, and standard deviations of the above indicators, by subject area and access type, are shown. It can be observed that means are higher than medians for all variables except for *References per Document* and *SJR Best Quartile*.

Finally, with respect to the statistical methods, we mainly used a non-parametric median test for the access type (OA vs. non-OA) in each area. Equality-of-median tests determine whether OA and non-OA journals are drawn from populations with the same median. Note that density functions for the variables analysed (except for *References per Document* and *SJR Best Quartile*) follow similar asymmetric shapes with long tails. For this reason, a contrast of medians was used instead of means.

**Results**

*Prevalence of the OA journals by subject areas*

Size and prevalence of gold OA, in the Scopus subject areas, are shown in Table 1. Note that some journals are assigned to two or more subject categories from different subject areas, so the total number of journals is less than the sum of columns 4 and 5. The largest subject areas (in the number of journals) in Scopus in year 2015 are: 'Medicine' (6,350), 'Social Sciences' (5,036), 'Arts and Humanities' (3,220), and 'Engineering' (2,267). In relation to the number of OA journals, the largest subject areas are again 'Medicine' (1,306) and 'Social Sciences' (713). However, the subject areas with a higher prevalence of OA in 2015 are: 'Multidisciplinary' (27.5%), 'Veterinary' (25.2%), 'Dentistry' (23.7%), 'Immunology and Microbiology' (23.5%), and 'Agricultural and Biological Sciences' (23.3%). By contrast, the subject areas in Scopus with a lower OA prevalence in 2015 are: 'Business, Management and Accounting' (6.6%), 'Economics, Econometrics and Finance' (10%), 'Decision Sciences' (10.8%), and 'Engineering' (11%).

[Table 1 about here]

*Differences between access types (OA vs. non-OA) in the amount of published and referenced documents in the journals*

Table 2 compares the median values of OA journals and non-OA ones by subject area, and shows the results of a median test for the variables *Total Documents* and *References per Document*. *Total Documents* is, in general, higher for non-OA journals. More specifically, in fourteen cases, non-OA journals present higher median values for total documents, rejecting the null hypothesis that journals in these areas are drawn from populations with comparable numbers of documents. Therefore, it can be concluded that in more than half of the subject areas (14 out of 27) the OA journals are smaller (significant at the 10% level) in terms of the number of documents published in 2015. In the rest of the cases (13 out of 27), there are no significant differences in size between access types.

[Table 2 about here]

Note that in the 'Multidisciplinary' subject area, the OA mega-journal *Scientific Reports* (10,867 documents in 2015) contrasts with non-OA journals like *PNAS* (3,857), *Nature* (2,653), and *Science* (2,106), all of which had over two thousand papers in the year 2015. Surprisingly, the OA mega-journal *PLoS ONE* is not found in the multidisciplinary subject area in the Scopus database in 2015.

However, *References per Document* shows a different behaviour to *Total Documents*. In general, the median numbers of references per document from OA and non-OA journals are quite similar in most of the cases. There are only 6 cases (out of 27) where the null hypothesis is rejected, that is, six cases in which the difference is statistically significant at the 5% level between the median numbers of referenced documents for OA journals as opposed to non-OA journals (5 of which are significant at the 1% level). In four cases, 'Medicine', 'Multidisciplinary', 'Nursing', and 'Earth and Planetary Sciences', OA journals have higher medians, whereas in the remaining two, 'Arts and Humanities' and 'Social Sciences', it is non-OA journals that have the higher median. Overall, this finding suggests that documents published in OA and non-OA journals are quite similar in terms of their propensity to reference other documents.

*Differences in the journal impact between access types (OA vs. non-OA)*

Table 3 compares the median behaviour of OA journals with non-OA ones by subject area for the variable *Cites per Document*. Additionally, a non-parametric median test is performed for each area. Thirteen of the twenty seven subject areas reject the null hypothesis of having the same median (for OA and non-OA types of journals) at 5% significance level for the variable *Cites per Document (3 years)*. That is, there is a significant difference in citation impact of journals (using a 3-year window) for 13 of 27 areas; of these 13, only two subject areas, 'Multidisciplinary' and 'Medicine', show a higher median for OA journals.

[Table 3 about here]

Similar conclusions are obtained for the variable *Cites per Document (2 years)*: sixteen of the twenty seven subject areas are found to come from populations with different medians for OA and non-OA journals (at 5% significance level). Again, of these, only 'Multidisciplinary' and 'Medicine' show a higher median for the OA journals.

Although there are some differences in relation to certain subject areas depending on the citation time window (2 and 3 years), in general they are not important and the conclusions are basically the same.

Although the early view postulate could explain the advantage of citation observed at paper level in the literature, this would not translate into advantage at journal level. One possible explication is that in those areas where anticipating pre-print or post-print provides a competitive citation advantage, paywalled journals often allow authors to deposit one of these two versions of the paper in repositories. According to SHERPA/RoMEO statistics (consulted on July 1, 2017), 80% of publishers formally allow some form of self-archiving, which may explain the small differences observed between the 2-year and 3-year citation windows.

Table 4 compares median values of OA journals and non-OA ones by subject area, and shows a median test for the variable *H Index*. This variable rejects the null hypothesis that both OA and non-OA journals are drawn from populations with the same median in 25 out of 27 cases at the 5% significance level. Only in the area of 'Arts and Humanities' is the median higher for OA journals. Therefore, there is a clear superiority of the non-OA type of journals for this variable. The smaller size and lower cites per document of OA journals may explain why these journals also have generally smaller *H Indexes*.

[Table 4 about here]

Figure 1 shows the density functions for the variable *SJR* by subject area and access type. Both the OA and non-OA density functions follow similar asymmetric shapes with long tails; consequently, in all the cases, the mode for both access types corresponds to approximately the same value of the *SJR* indicator. In most cases, the maximum density for OA is higher than that for non-OA.

[Figure 1 about here]

This type of highly biased distribution indicates that, for both access types, a majority of the journals present relatively small values of the *SJR* indicator. The two distributions, according to access type, intercept at low values. As a general rule, the distribution of OA begins above non-OA for the smaller values of the citation impact indicator and continues below for intermediate values. Only in a few subject areas is this trend reversed. This is the case of 'Medicine', 'Multidisciplinary', and 'Neuroscience'.

The tail of the distribution collects journals in which the indicator takes the highest values. Thus, in only 4 subject areas is the tail of the OA distribution visibly longer than that of the non-OA. However, in most cases the opposite occurs. This result suggests again that, in general, OA journals have lower impacts than non-OA ones.

Table 5 compares the median behaviour of OA journals with non-OA ones by subject area for the variable *SJR*. Additionally, a non-parametric median test is performed for each area. Eighteen of the twenty seven subject areas reject the null hypothesis that OA and non-OA journals are drawn from populations with the same median (16 of which are at 5% significance level), and in all these cases medians from non-OA journals are higher than those from OA ones.

[Table 5 about here]

Therefore, in no subject area is there an OA citation advantage. This poor result could be explained by the difference in the quartile to which the journals belong according to access type, and the age of the journal.

*OA citation advantage controlling by the quartile of the journal*

All journals indexed in the Scopus database are classified into one or several subject categories, and later each category is assigned to a research area. The SJR data of a journal is related to different citation habits within each category. Thus, it is wrong to make comparisons of SJR figures in journals belonging to different categories or areas. To carry out this type of comparison sorting into quartiles was used. With this indicator we were able to compare journals in different categories without considering the difference in the SJR figures.

Quartiles are the values that divide the set of journals arranged in four equal percentage parts. So, Q1 denotes the top 25% of the SJR distribution, Q2 a mid-high position (between top 50% and top 25%), Q3 mid-low position (top 75% to top 50%), and Q4 the lowest position (bottom 25% of the SJR distribution). Given that each journal belongs to one or several categories, the best quartile is defined as the top of all the quartiles in the categories in which the journal is indexed.

In this sense, Figure 2 shows the percentages of journals included in *SJR* quartiles, by subject area and access type. These graphs evidence once again the superiority, in general, of non-OA journals with respect to OA ones in the year 2015. In this regard, the percentage of journals included in the first quartile is higher for the non-OA group in 24 subject areas (out of 27). More specifically, in sixteen cases, the percentage of journals included in the first quartile is more than ten points higher for the non-OA group. Nevertheless, this clear superiority of non-OA journals in the first quartile is offset in the second and third quartiles. Finally, the number of areas with a higher percentage of journals in the fourth quartile is similar for both types of access. Therefore, this fact may partly explain the poor results obtained by OA journals in relation to citation impact.

[Figure 2 about here]

Finally, Table 6 compares the median values of the variable *SJR* considering four alternative subsamples depending on the best quartile that characterizes each journal. In most areas, there are no significant differences at the 10% level between the OA and non-OA median of the variable *SJR*. This result holds for all four subgroups considered in the analysis (Q1 to Q4).

[Table 6 about here]

More specifically, in the Q1 subgroup, significant median test results at the 10% level were found in six areas: five of them in favour of non-OA journals ('Social Sciences', 'Pharmacology, Toxicology and Pharmaceutics', 'Business, Management and Accounting', 'Arts and Humanities', and 'Computer Science') and only one in favour of OA journals ('Physics and Astronomy'). In subgroups Q2 and Q3, six significant median test results at the 10% level were found in favour of non-OA journals, and only in subgroup Q3 was a significant median test result at the 10% level found in favour of OA journals for two areas. Finally, subgroup Q4 differs from the other subgroups, since all nine significant median test results at the 10% level were in favour of OA journals.

That is, there is an OA citation advantage (at a significance level of 10%) in 9 research areas in the subgroup of journals with lower impact.

*OA citation advantage controlling by the age of the journal*

In general, it is assumed that older journals are more prestigious and so better cited. In order to control for this, we proxied the age of the journal by the first year the journal was indexed in the Scopus database. In this way, Table 7 considers only young journals, specifically those who were first included in the Scopus database from 2005 to 2014. *SJR* indicator values were compared between OA and non-OA journals, by areas. Additionally, this ten year sample was divided into two subsamples (2005-2009 and 2010-2014) in order to detect whether or not there were significant differences in the comparison of OA vs. non-OA journals with respect to the *SJR* indicator between the two periods.

It can be seen from Table 7 that, in general, there are no major differences between journals which first entered the Scopus database in the period 2005-2009 and 2010-2014. However, when comparing results from Table 7 with Table 5, where all journals were considered, it can be concluded that journal age matters in the analysis of *SJR* indicators. Specifically, when all journals are considered, in 18 out of 27 areas non-OA journals show higher *SJR* indicators than OA ones, and OA journals were not found to be superior in any case. On the contrary, when considering young journals (period 2005-2014), little evidence is found in favour of OA or non-OA journals, since in most of the areas there are no significant differences between the two journal types. For the period 2010-2014, the *SJR* indicator is superior for non-OA journals in two areas ('Psychology' and 'Business, Management and Accounting'), both significant at the 1% level, and is found to be superior for OA journals in four areas ('Medicine', 'Pharmacology, Toxicology and Pharmaceutics', 'Computer Science', and 'Engineering'), two of which are significant at the 1% level. Therefore, it seems that the gap between open and non-open access journals decreases over the years when the analysis is controlled by journal age.

**Discussion**

Our focus on journal-level impact (rather than paper-level impact) refutes the *OA Citation Advantage postulate*. This result corroborates in part the conclusion of Archambault et al. (2014) for a large data analysis. In a sample of one and a half million Scopus indexed papers published between 2007 and 2012, they estimated that more than 50% of the scientific papers could be downloaded for free on the Internet. Furthermore, in a set of about half a million papers published between 2009 and 2011, the OA papers (about 250,000) had a 26% citation advantage compared to the full set of about 500,000 papers. However, the advantage varied by type of OA: green (author posting) had a 53% advantage, while gold (journal OA) had a 39% disadvantage (i.e. they had a lower citation rate than the full set). The advantage also varied by field, with the highest advantage seen in some humanities fields and the lowest in medical fields.

Analysing article downloads -usage- is a complementary method for studying the effects of OA. Davis et al. (2008), for articles in psychology journals published by one publisher, concluded that OA articles were downloaded more often than papers with subscription-based access. However, they did not find a significant effect of open access on citations.

Consequently, the citation advantage in pros of OA observed by other authors in the literature should be justified by the *Early View and Selection Bias postulates*. Thus, Kurtz et al. (2005) found in a study on astronomy articles evidence of a *selection bias* (authors post their best articles freely on the web) and an *early view effect* (articles deposited as pre-prints are consulted earlier and are therefore cited more often). As a confirmation, Moed (2007) found that for physics articles these two effects may explain a large part of the differences in citation impact between journal articles posted as pre-print in *ArXiv* (a pre-print server hosted by Cornell University) and papers that were not.

Our study suggests that there is no OA citation advantage at the journal level. The early view postulate would explain the advantage of citation observed at paper level. However, this would not translate into advantage at journal level. This is because in those areas where anticipating pre-print or post-print provides a competitive advantage, paywalled journals often allow authors to deposit one of these two versions of the paper in repositories, so the advantage of OA at the journal level is reduced. According to the SHERPA/RoMEO database of publishers' policies on copyright and self-archiving, 80% of publishers formally allow some form of self-archiving (statistics for the 2380 publishers in the database (consulted on July 1, 2017)).

Therefore, at journal level one could speak of citation advantage related to reputation as a combination of paper average impact, journal visibility and access, regardless of how this access occurs.

**Conclusions**

The present paper takes a journal-level approach to assessing the OA citation advantage, whereas many others take a paper-level approach. This is the novel contribution of the paper.

The results for the analysis of gold OA in the 27 subject areas of the Scopus database are presented. The data were obtained from Scimago, sorted using the Scopus database and tagged as OA/non-OA using the DOAJ list. For a total of 3,737 OA journals (16.8%) and 18,485 non-OA journals (83.2%), no generalizable gold OA citation advantage is found at journal level. In fact, in many areas a gold OA citation disadvantage is observed.

With respect to OA prevalence in the Scopus database, the highest percentages of OA journals are over 25% ('Multidisciplinary' and 'Veterinary') and the lowest are below 10% ('Business, Management and Accounting' and 'Economics, Econometrics and Finance').

As to the differences in production and referencing in the year 2015, the OA journals are smaller, in terms of the number of documents published, in more than half of the subject areas. However, in general there are no significant differences (below 5%) between groups of journals in the amount of referenced documents. Thus, it can be concluded that document typology published in the OA journals is quite similar to that of non-OA ones.

Cites per document are similar with respect to the citation time window, and half of the subject areas are found to come from populations with different medians depending on the type of access. Only 'Multidisciplinary' and 'Medicine' show higher medians for the OA journals. Moreover, smaller sizes and lower cites per document of OA journals may explain the clear superiority of non-OA journals for the *H Index*.

In 18 of the 27 subject areas, *SJR* medians from non-OA journals are significantly higher at the 10% level than those from OA ones (12 of them at the 1% level). Moreover, when we consider only the journals of greatest impact (first quartile), the advantage of OA is observed in only one of the 27 areas, whereas the opposite happens (advantage for non-OA) in 5 of the 27 areas. Therefore, there is no generalizable OA citation advantage.

Some considerations can be made with respect to these unexpected results, most importantly concerning journal visibility. Most OA journals are not at the top of rankings that measure the impact of the journals (e.g. first quartile in the *SJR*). However, the top of these rankings provides high visibility for the journals, at the same time as access through subscription is generalized for these top ranked journals in most research institutions. Thus, gold OA does not guarantee higher visibility in relation to the subscription model.

Importantly, we do not take into account the influence of APC costs. The APC of top ranked journals is evidently higher than that of lower ranked ones. For this reason, many authors cannot publish in some desired gold OA journals, especially in the top ranked ones.

About the limitations of our study, the following bias must be underlined. It is based on citation analyses carried out for a database with a selective coverage, in most research areas, of international top journals in the disciplines instead of lower impact or more nationally oriented journals. This occurs in all research areas in the Scopus database but more strongly for 'Arts and Humanities', 'Business, Management and Accounting', 'Economics, Econometrics and Finance', and 'Social Sciences'. Authors who publish in such kinds of international top journal – a necessary condition for citations to be recorded in the study – will tend to have access to these journals anyway. That is, there may be a positive effect of OA upon citation impact, but it is not visible in the database used in some areas.

About the reasons why there is no OA advantage or even OA disadvantage at journal level, we believe that is the personal motivation of each author individually that leads him/her to try to improve the visibility and access of its works. For this purpose, it uses the green access as well as comments on blogs and social networks. For these authors, there is an advantage of citation at the paper level. However, by adding the impact of many authors in a journal, the fact that it is an open access journal makes the responsibility for dissemination diluted between the publisher and the group of authors, causing some of them to consider that the task of dissemination must rest exclusively on the journal. In this way, subscription journals benefit from its subscriber network and at the same time from the effort of many of its authors to facilitate access through the greenway.

In relation to the researcher's or institution's decision about impact-maximizing strategies, a focus on journal-level impact (rather than paper-level impact) connects to the idea that the final decision is independent of the business model of the journal.

The average impact of a journal is not generally related to the business model used by the publisher. There are high impact journals both OA and paywalled, and opting for a unique business model is not an optimal decision. In each area and at any time, authors and subscribers should choose to publish and maintain the subscription of the journals of greatest impact in the specific specialty. That is, the researchers cite from among the papers to which they have access, regardless of whether such access is produced because it is free or the subscription is available, those considered more relevant.

Therefore, and from our point of view, an optimal citation strategy for authors is to publish in the journals of their discipline of greatest impact, regardless of the type of access, and at the same time make use of the thematic and institutional repositories. In the institutional case, an optimal strategy for managers is to maintain the subscription to the journals of greater impact in each discipline and to dedicate the money that was destined to the subscription of those of smaller impact to pay for the costs of publication in the OA journals with greater impact in each discipline.

**Acknowledgements**

This research has been supported by the Ministerio de Economía, Industria y Competitividad of Spain under the research project ECO2014-59067-P.

Table 1: Size and prevalence of gold OA in 2015 for the Scopus subject areas

|   | Scopus Subject Areas | Abbreviation | Total Journals | OA Journals | OA Prevalence |
|---|---|---|---|---|---|
| 1 | Multidisciplinary | MULT | 109 | 30 | 27.5% |
| 2 | Veterinary | VETE | 214 | 54 | 25.2% |
| 3 | Dentistry | DENT | 169 | 40 | 23.7% |
| 4 | Immunology and Microbiology | IMMU | 510 | 120 | 23.5% |
| 5 | Agricultural and Biological Sciences | AGRI | 1,860 | 434 | 23.3% |
| 6 | Biochemistry, Genetics and Molecular Biology | BIOC | 1,848 | 418 | 22.6% |
| 7 | Neuroscience | NEUR | 499 | 108 | 21.6% |
| 8 | Medicine | MEDI | 6,350 | 1,306 | 20.6% |
| 9 | Pharmacology, Toxicology and Pharmaceutics | PHAR | 709 | 142 | 20.0% |
| 10 | Environmental Science | ENVI | 1,152 | 204 | 17.7% |
| 11 | Earth and Planetary Sciences | EART | 1,018 | 173 | 17.0% |
| 12 | Chemical Engineering | CENG | 494 | 79 | 16.0% |
| 13 | Computer Science | COMP | 1,316 | 193 | 14.7% |
| 14 | Chemistry | CHEM | 763 | 110 | 14.4% |
| 15 | Social Sciences | SOCI | 5,036 | 713 | 14.2% |
| 16 | Health Professions | HEAL | 460 | 65 | 14.1% |
| 17 | Mathematics | MATH | 1,252 | 170 | 13.6% |
| 18 | Physics and Astronomy | PHYS | 953 | 125 | 13.1% |
| 19 | Materials Science | MATE | 999 | 120 | 12.0% |
| 20 | Psychology | PSYC | 1,044 | 123 | 11.8% |
| 21 | Energy | ENER | 341 | 40 | 11.7% |
| 22 | Arts and Humanities | ARTS | 3,220 | 362 | 11.2% |
| 23 | Nursing | NURS | 571 | 64 | 11.2% |
| 24 | Engineering | ENGI | 2,267 | 250 | 11.0% |
| 25 | Decision Sciences | DECI | 297 | 32 | 10.8% |
| 26 | Economics, Econometrics and Finance | ECON | 817 | 82 | 10.0% |
| 27 | Business, Management and Accounting | BUSI | 1,081 | 71 | 6.6% |
|   | Total of journals |   | 22,222* | 3,737* | 16.8% |

* Journals are assigned to one or several subject categories from different areas, so the total of journals is less than the sum of the column.

OA Prevalence in [25, 30)
OA Prevalence in [20, 25)
OA Prevalence in [15, 20)
OA Prevalence in [10, 15)
OA Prevalence in [5, 10)

**Table 2:** Median values by access type, and equality of median tests for the variables *Total Documents* and *References per Document* (year 2015)

| | Total Documents | | | | References per Document | | |
|---|---|---|---|---|---|---|---|
| Areas | Median OA | Median Non-OA | Median test p-value | Areas | Median OA | Median Non-OA | Median test p-value |
| MULT | 93 | 42 | 0.480 | MEDI | 26.0 | 22.7 | 0.000 |
| HEAL | 49 | 49 | 0.940 | MULT | 26.5 | 18.9 | 0.004 |
| VETE | 49 | 54 | 0.938 | NURS | 27.9 | 22.3 | 0.005 |
| ARTS | 18 | 18 | 0.880 | EART | 40.2 | 33.4 | 0.030 |
| NEUR | 59.5 | 78 | 0.490 | ENGI | 23.7 | 22.5 | 0.150 |
| ENER | 59.5 | 71 | 0.410 | MATH | 23.6 | 22.2 | 0.160 |
| DENT | 44.5 | 56 | 0.221 | DENT | 25.7 | 23.2 | 0.190 |
| ECON | 23 | 27 | 0.213 | COMP | 30.3 | 27.9 | 0.210 |
| ENVI | 38 | 43 | 0.212 | ENER | 27.5 | 23.3 | 0.230 |
| MEDI | 55 | 58 | 0.200 | NEUR | 42.9 | 39.0 | 0.310 |
| BUSI | 24 | 29 | 0.160 | IMMU | 36.6 | 35.5 | 0.340 |
| DECI | 24.5 | 37 | 0.130 | PSYC | 38.9 | 38.4 | 0.440 |
| AGRI | 44 | 49 | 0.130 | ENVI | 36.9 | 35.4 | 0.580 |
| NURS | 33 | 54 | 0.087 | HEAL | 28.4 | 28.0 | 0.780 |
| IMMU | 60.5 | 83 | 0.084 | CENG | 30.8 | 30.0 | 0.800 |
| PHAR | 47 | 69 | 0.066 | DECI | 31.0 | 30.9 | 0.860 |
| SOCI | 22 | 24 | 0.055 | ECON | 34.7 | 34.1 | 0.880 |
| COMP | 33 | 42 | 0.029 | MATE | 26.3 | 25.9 | 0.910 |
| MATH | 37 | 48.5 | 0.017 | PHAR | 32.4 | 31.7 | 0.910 |
| MATE | 62 | 96 | 0.012 | PHYS | 27.6 | 27.4 | 0.990 |
| ENGI | 44 | 63 | 0.007 | VETE | 28.3 | 26.8 | 0.630 |
| CENG | 63 | 102 | 0.001 | CHEM | 31.5 | 32.4 | 0.617 |
| EART | 27 | 43 | 0.001 | AGRI | 34.7 | 35.5 | 0.470 |
| PSYC | 25 | 37 | 0.001 | BUSI | 39.6 | 42.0 | 0.470 |
| BIOC | 53 | 91 | 0.000 | BIOC | 36.3 | 36.9 | 0.340 |
| CHEM | 73.5 | 135 | 0.000 | SOCI | 32.6 | 36.4 | 0.000 |
| PHYS | 67 | 115 | 0.000 | ARTS | 29.5 | 32.9 | 0.000 |

**Note:** Equality of median test determines whether OA and non-OA journals are drawn from populations with the same median.

| Equality of medians | | |
|---|---|---|
| | Me(OA)>=Me(non-OA) | Significant at 1% |
| | Me(OA)>=Me(non-OA) | Significant at 5% |
| | Me(OA)>=Me(non-OA) | Significant at 10% |
| | Me(OA)>=Me(non-OA) | Non-significant |
| | Me(OA)<Me(non-OA) | Non-significant |
| | Me(OA)<Me(non-OA) | Significant at 10% |
| | Me(OA)<Me(non-OA) | Significant at 5% |
| | Me(OA)<Me(non-OA) | Significant at 1% |

**Table 3:** Median values by access type, and equality of median tests for the variable *Cites per Document*

| Areas | Cites per Document (3 years) | | | Cites per Document (2 years) | | |
|---|---|---|---|---|---|---|
| | Median OA | Median Non-OA | Median test p-value | Median OA | Median Non-OA | Median test p-value |
| MULT | 0.484 | 0.265 | 0.008 | 0.450 | 0.250 | 0.008 |
| MEDI | 0.903 | 0.805 | 0.038 | 0.990 | 0.860 | 0.022 |
| NEUR | 2.121 | 1.980 | 0.433 | 2.030 | 2.020 | 0.932 |
| NURS | 0.577 | 0.622 | 0.906 | 0.540 | 0.640 | 0.536 |
| ENVI | 0.902 | 0.949 | 0.700 | 0.885 | 0.940 | 0.817 |
| ENER | 0.654 | 0.724 | 0.628 | 0.610 | 0.840 | 0.247 |
| HEAL | 0.755 | 0.877 | 0.592 | 0.680 | 0.920 | 0.619 |
| VETE | 0.465 | 0.533 | 0.431 | 0.425 | 0.535 | 0.157 |
| PHAR | 1.086 | 1.267 | 0.311 | 1.075 | 1.350 | 0.230 |
| IMMU | 1.609 | 1.900 | 0.251 | 1.515 | 1.905 | 0.053 |
| DENT | 0.575 | 0.932 | 0.221 | 0.565 | 0.970 | 0.380 |
| ENGI | 0.610 | 0.697 | 0.205 | 0.560 | 0.700 | 0.036 |
| EART | 0.685 | 0.881 | 0.133 | 0.700 | 0.880 | 0.120 |
| COMP | 0.809 | 0.982 | 0.119 | 0.720 | 0.980 | 0.008 |
| MATE | 0.784 | 0.956 | 0.100 | 0.760 | 0.970 | 0.014 |
| MATH | 0.571 | 0.700 | 0.058 | 0.525 | 0.700 | 0.002 |
| CENG | 0.843 | 1.238 | 0.050 | 0.770 | 1.250 | 0.027 |
| DECI | 0.701 | 0.946 | 0.042 | 0.580 | 0.900 | 0.009 |
| BIOC | 1.619 | 1.898 | 0.017 | 1.625 | 1.940 | 0.005 |
| PHYS | 0.731 | 1.019 | 0.001 | 0.740 | 1.050 | 0.002 |
| AGRI | 0.660 | 0.891 | 0.000 | 0.625 | 0.870 | 0.000 |
| ARTS | 0.086 | 0.149 | 0.000 | 0.080 | 0.130 | 0.000 |
| BUSI | 0.455 | 0.774 | 0.000 | 0.390 | 0.735 | 0.000 |
| CHEM | 0.783 | 1.502 | 0.000 | 0.730 | 1.510 | 0.000 |
| ECON | 0.314 | 0.634 | 0.000 | 0.265 | 0.600 | 0.000 |
| PSYC | 0.492 | 1.072 | 0.000 | 0.450 | 1.030 | 0.000 |
| SOCI | 0.222 | 0.452 | 0.000 | 0.190 | 0.430 | 0.000 |

**Note:** Equality of median test determines whether OA and non-OA journals are drawn from populations with the same median.

| Equality of medians | | |
|---|---|---|
| | Me(OA)>=Me(non-OA) | Significant at 1% |
| | Me(OA)>=Me(non-OA) | Significant at 5% |
| | Me(OA)>=Me(non-OA) | Significant at 10% |
| | Me(OA)>=Me(non-OA) | Non-significant |
| | Me(OA)<Me(non-OA) | Non-significant |
| | Me(OA)<Me(non-OA) | Significant at 10% |
| | Me(OA)<Me(non-OA) | Significant at 5% |
| | Me(OA)<Me(non-OA) | Significant at 1% |

**Table 4:** Median values by access type, and equality of median tests for the variable *H Index* in the year 2015

| Areas | H Index Median OA | Median Non-OA | Median test p-value |
|---|---|---|---|
| ARTS | 5 | 3 | 0.000 |
| MULT | 15 | 13 | 0.610 |
| ENER | 11 | 16 | 0.180 |
| VETE | 12.5 | 21 | 0.023 |
| DECI | 11.5 | 21 | 0.016 |
| ENVI | 14 | 20 | 0.004 |
| NURS | 11 | 20 | 0.003 |
| CENG | 14 | 28 | 0.001 |
| DENT | 8 | 29 | 0.001 |
| EART | 13 | 19 | 0.001 |
| AGRI | 12 | 24 | 0.000 |
| BIOC | 20 | 48 | 0.000 |
| BUSI | 7 | 14 | 0.000 |
| CHEM | 16.5 | 37 | 0.000 |
| COMP | 12 | 22 | 0.000 |
| ECON | 6 | 14 | 0.000 |
| ENGI | 10 | 17 | 0.000 |
| HEAL | 11 | 23 | 0.000 |
| IMMU | 19.5 | 46 | 0.000 |
| MATE | 15 | 25 | 0.000 |
| MATH | 12 | 22 | 0.000 |
| MEDI | 14 | 21 | 0.000 |
| NEUR | 22.5 | 51 | 0.000 |
| PHAR | 15 | 28 | 0.000 |
| PHYS | 17 | 33 | 0.000 |
| PSYC | 9 | 28 | 0.000 |
| SOCI | 5 | 10 | 0.000 |

**Note:** Equality of median test determines whether OA and non-OA journals are drawn from populations with the same median.

| Equality of medians | | |
|---|---|---|
| | Me(OA)>=Me(non-OA) | Significant at 1% |
| | Me(OA)>=Me(non-OA) | Significant at 5% |
| | Me(OA)>=Me(non-OA) | Significant at 10% |
| | Me(OA)>=Me(non-OA) | Non-significant |
| | Me(OA)<Me(non-OA) | Non-significant |
| | Me(OA)<Me(non-OA) | Significant at 10% |
| | Me(OA)<Me(non-OA) | Significant at 5% |
| | Me(OA)<Me(non-OA) | Significant at 1% |

**Figure 1:** Density functions of OA journals (long dash-dot black line) and non-OA journals (solid red line) for the variable SJR. The x axis corresponds to the value of the SJR indicator, and the y axis to the density of journals in the research area with that value of the SJR indicator.

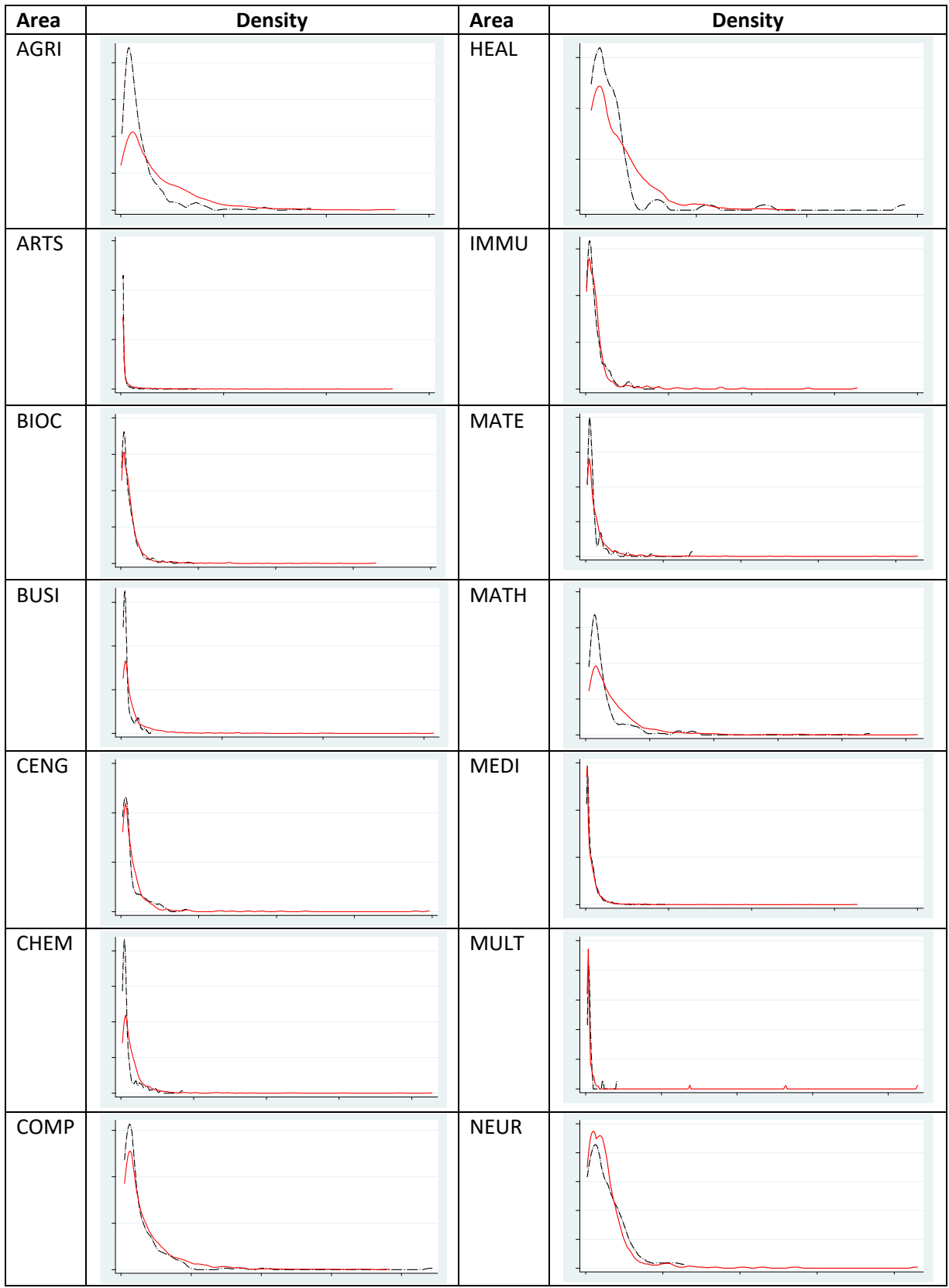

**Figure 1:** (continuation)

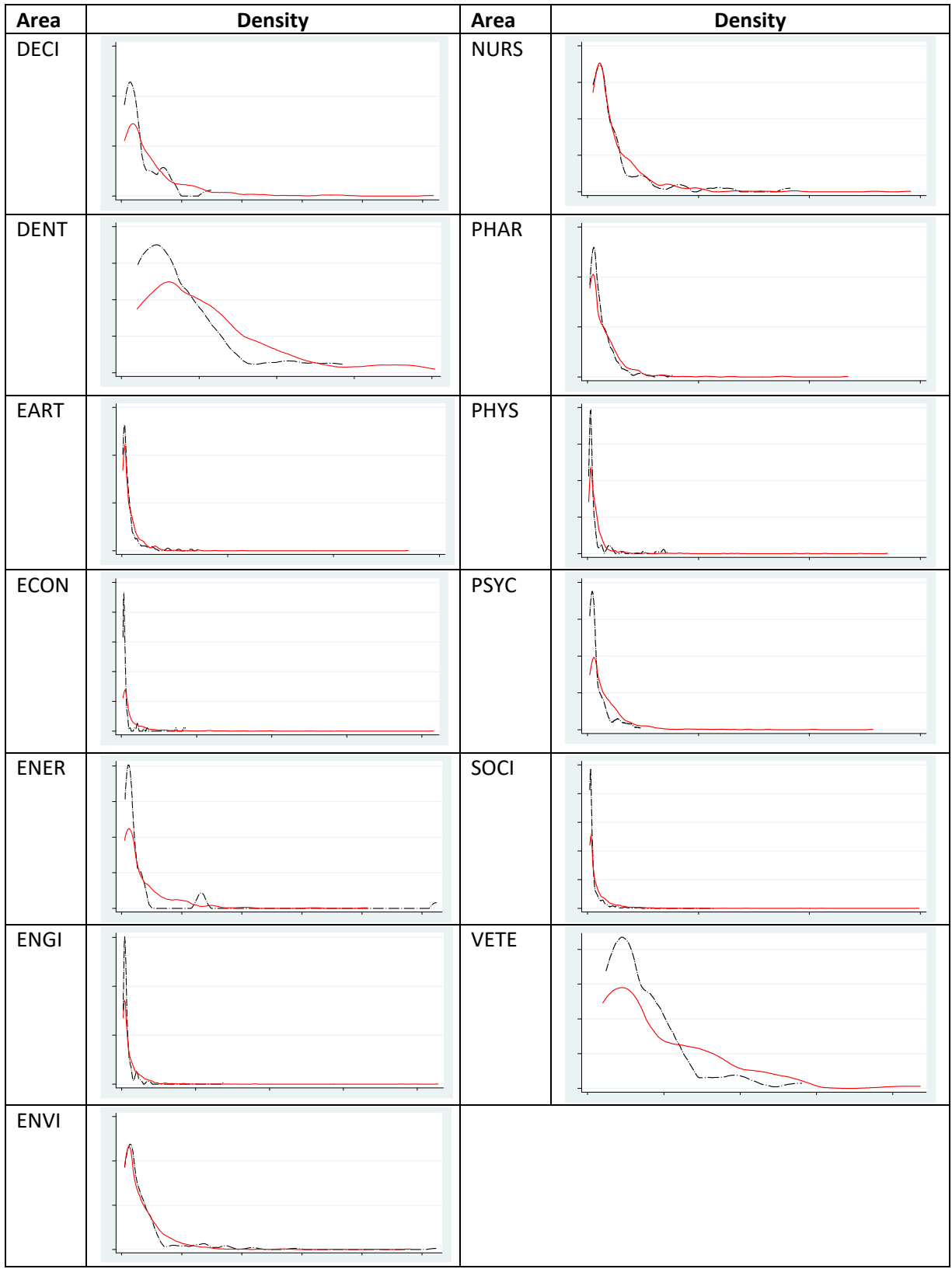

**Table 5:** Median values by access type, and equality of median tests for the variable *SJR*

| Areas | SJR | | |
|---|---|---|---|
| | Median | Median | Median test |
| | OA | Non-OA | p-value |
| MULT | 0.224 | 0.191 | 0.258 |
| MEDI | 0.401 | 0.394 | 0.798 |
| ENVI | 0.445 | 0.456 | 0.938 |
| NURS | 0.265 | 0.322 | 0.930 |
| VETE | 0.286 | 0.301 | 0.875 |
| NEUR | 0.982 | 1.028 | 0.762 |
| HEAL | 0.354 | 0.431 | 0.284 |
| EART | 0.376 | 0.460 | 0.243 |
| IMMU | 0.744 | 0.919 | 0.175 |
| DECI | 0.344 | 0.657 | 0.096 |
| DENT | 0.265 | 0.455 | 0.051 |
| ENER | 0.282 | 0.384 | 0.037 |
| PHAR | 0.393 | 0.468 | 0.032 |
| CENG | 0.340 | 0.474 | 0.014 |
| BIOC | 0.702 | 0.841 | 0.012 |
| MATE | 0.342 | 0.413 | 0.010 |
| COMP | 0.324 | 0.446 | 0.008 |
| ARTS | 0.113 | 0.129 | 0.003 |
| ENGI | 0.257 | 0.342 | 0.000 |
| MATH | 0.360 | 0.600 | 0.000 |
| PHYS | 0.327 | 0.515 | 0.000 |
| AGRI | 0.345 | 0.460 | 0.000 |
| BUSI | 0.205 | 0.352 | 0.000 |
| CHEM | 0.282 | 0.526 | 0.000 |
| ECON | 0.189 | 0.354 | 0.000 |
| PSYC | 0.232 | 0.536 | 0.000 |
| SOCI | 0.168 | 0.261 | 0.000 |

**Note:** Equality of median test determines whether OA and non-OA journals are drawn from populations with the same median.

| Equality of medians | |
|---|---|
| Me(OA)>=Me(non-OA) | Significant at 1% |
| Me(OA)>=Me(non-OA) | Significant at 5% |
| Me(OA)>=Me(non-OA) | Significant at 10% |
| Me(OA)>=Me(non-OA) | Non-significant |
| Me(OA)<Me(non-OA) | Non-significant |
| Me(OA)<Me(non-OA) | Significant at 10% |
| Me(OA)<Me(non-OA) | Significant at 5% |
| Me(OA)<Me(non-OA) | Significant at 1% |

**Figure 2:** Percentage of journals whose *SJR Best Quartile* is Q1, Q2, Q3, and Q4, by area and access type

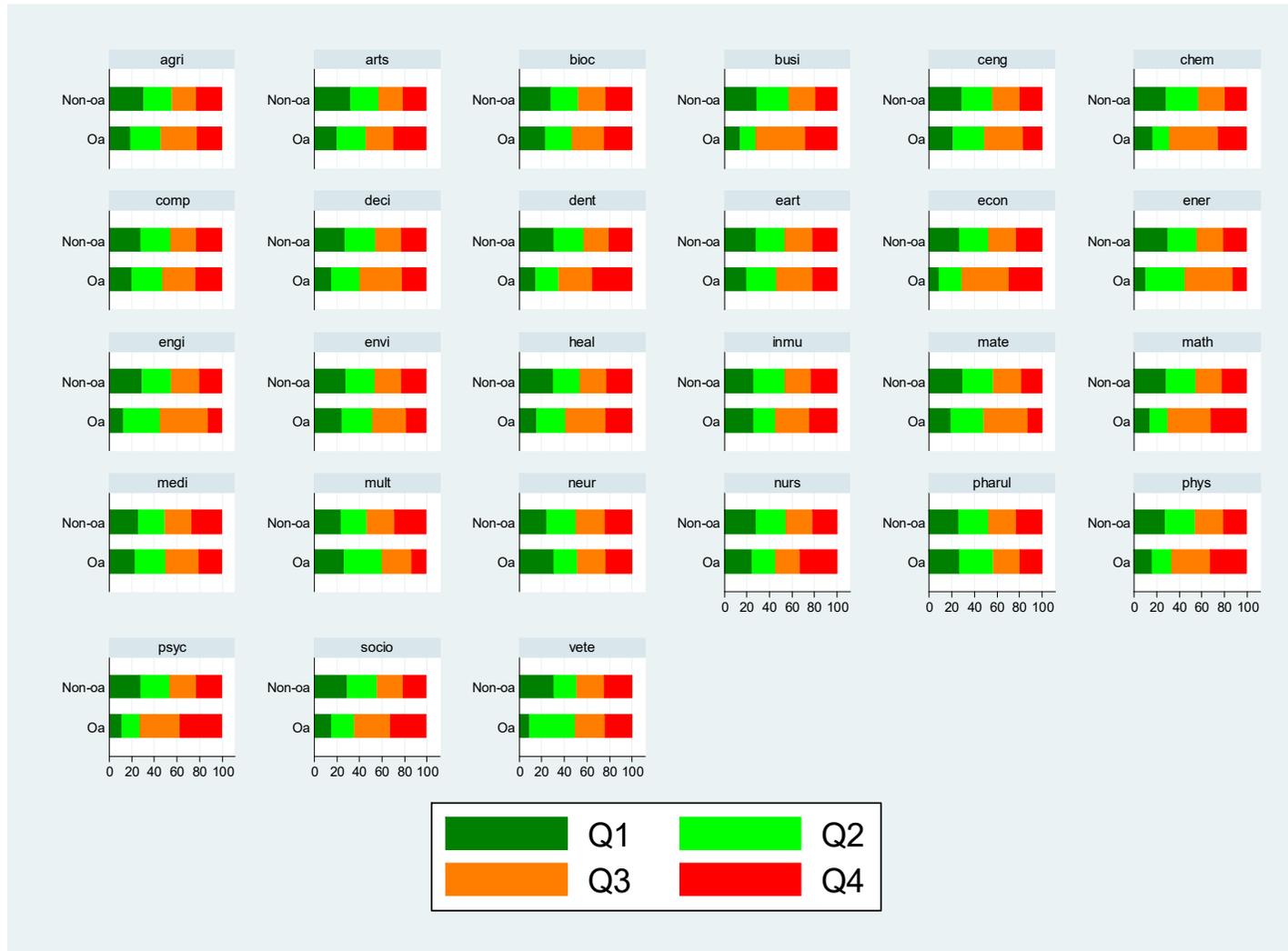

**Table 6:** Median values by access type, and equality of median tests for the variable *SJR* (by area and quartile)

| | Q1 | | | | Q2 | | | | Q3 | | | | Q4 | | |
|---|---|---|---|---|---|---|---|---|---|---|---|---|---|---|---|
| Areas | Median | Median | Median test | Areas | Median | Median | Median test | Areas | Median | Median | Median test | Areas | Median | Median | Median test |
| | OA | Non-OA | p-value | | OA | Non-OA | p-value | | OA | Non-OA | p-value | | OA | Non-OA | p-value |
| PHYS | 2.05 | 1.31 | 0.04 | MULT | 0.24 | 0.21 | 0.24 | MEDI | 0.27 | 0.24 | 0.00 | MEDI | 0.12 | 0.11 | 0.00 |
| ENER | 2.69 | 1.45 | 0.12 | MEDI | 0.62 | 0.60 | 0.41 | ARTS | 0.11 | 0.10 | 0.05 | VETE | 0.14 | 0.12 | 0.00 |
| CENG | 1.74 | 1.26 | 0.29 | IMMU | 1.14 | 1.07 | 0.42 | MATE | 0.25 | 0.23 | 0.52 | BIOC | 0.19 | 0.16 | 0.01 |
| VETE | 0.95 | 0.84 | 0.35 | NURS | 0.42 | 0.39 | 0.56 | SOCI | 0.18 | 0.17 | 0.56 | AGRI | 0.17 | 0.14 | 0.01 |
| MATH | 1.66 | 1.49 | 0.39 | DENT | 0.50 | 0.49 | 0.69 | PHAR | 0.27 | 0.25 | 0.70 | ENVI | 0.13 | 0.12 | 0.05 |
| DENT | 0.99 | 0.88 | 0.66 | ARTS | 0.15 | 0.14 | 0.74 | MULT | 0.17 | 0.17 | 0.96 | NURS | 0.13 | 0.11 | 0.05 |
| IMMU | 2.51 | 2.18 | 0.68 | ENER | 0.47 | 0.47 | 0.77 | HEAL | 0.29 | 0.28 | 1.00 | ENER | 0.13 | 0.11 | 0.06 |
| CHEM | 1.54 | 1.43 | 0.79 | EART | 0.61 | 0.60 | 0.81 | ENER | 0.22 | 0.22 | 0.96 | PSYC | 0.15 | 0.14 | 0.06 |
| AGRI | 1.24 | 1.18 | 0.79 | VETE | 0.43 | 0.43 | 0.96 | DENT | 0.24 | 0.26 | 0.81 | IMMU | 0.20 | 0.16 | 0.09 |
| PSYC | 1.53 | 1.51 | 0.80 | NEUR | 1.26 | 1.24 | 1.00 | NURS | 0.20 | 0.22 | 0.78 | ENGI | 0.12 | 0.11 | 0.11 |
| EART | 1.59 | 1.54 | 0.84 | BIOC | 1.06 | 1.06 | 0.93 | NEUR | 0.64 | 0.70 | 0.73 | MATH | 0.18 | 0.17 | 0.28 |
| NEUR | 2.46 | 2.28 | 0.96 | CENG | 0.51 | 0.53 | 0.85 | BUSI | 0.20 | 0.21 | 0.72 | PHYS | 0.18 | 0.18 | 0.39 |
| ECON | 1.71 | 1.51 | 0.99 | MATH | 0.69 | 0.74 | 0.84 | EART | 0.29 | 0.30 | 0.65 | BUSI | 0.13 | 0.11 | 0.44 |
| MATE | 1.26 | 1.24 | 1.00 | BUSI | 0.38 | 0.42 | 0.76 | AGRI | 0.28 | 0.29 | 0.56 | CENG | 0.18 | 0.12 | 0.47 |
| BIOC | 2.25 | 2.27 | 0.91 | PSYC | 0.67 | 0.73 | 0.65 | ENGI | 0.22 | 0.22 | 0.54 | PHAR | 0.12 | 0.11 | 0.53 |
| ENVI | 1.14 | 1.30 | 0.90 | ENVI | 0.56 | 0.58 | 0.57 | VETE | 0.20 | 0.22 | 0.49 | MATE | 0.12 | 0.12 | 0.57 |
| NURS | 0.75 | 0.77 | 0.81 | DECI | 0.77 | 0.84 | 0.31 | MATH | 0.37 | 0.39 | 0.42 | EART | 0.13 | 0.12 | 0.72 |
| HEAL | 1.01 | 1.06 | 0.76 | HEAL | 0.47 | 0.54 | 0.31 | ENVI | 0.24 | 0.27 | 0.25 | DENT | 0.12 | 0.12 | 0.92 |
| MEDI | 1.39 | 1.44 | 0.41 | PHYS | 0.57 | 0.66 | 0.27 | IMMU | 0.56 | 0.61 | 0.19 | ECON | 0.12 | 0.12 | 0.93 |
| DECI | 1.55 | 2.13 | 0.37 | CHEM | 0.55 | 0.63 | 0.19 | PHYS | 0.33 | 0.34 | 0.14 | NEUR | 0.18 | 0.18 | 1.00 |
| MULT | 0.36 | 0.46 | 0.25 | COMP | 0.49 | 0.58 | 0.14 | DECI | 0.28 | 0.39 | 0.11 | HEAL | 0.12 | 0.12 | 1.00 |
| ENGI | 1.04 | 1.16 | 0.20 | SOCI | 0.29 | 0.34 | 0.08 | CENG | 0.22 | 0.26 | 0.09 | DECI | 0.15 | 0.16 | 0.97 |
| COMP | 1.26 | 1.38 | 0.09 | MATE | 0.40 | 0.49 | 0.07 | COMP | 0.25 | 0.29 | 0.08 | SOCI | 0.10 | 0.10 | 0.79 |

| ARTS | 0.32 | 0.40 | 0.07 | PHAR | 0.42 | 0.71 | 0.04 | PSYC | 0.27 | 0.33 | 0.07 | MULT | 0.11 | 0.12 | 0.64 |
|---|---|---|---|---|---|---|---|---|---|---|---|---|---|---|---|
| BUSI | 0.85 | 1.30 | 0.02 | ECON | 0.27 | 0.52 | 0.01 | BIOC | 0.50 | 0.60 | 0.03 | CHEM | 0.14 | 0.16 | 0.33 |
| PHAR | 1.06 | 1.37 | 0.02 | ENGI | 0.37 | 0.43 | 0.01 | ECON | 0.19 | 0.24 | 0.01 | ARTS | 0.10 | 0.10 | 0.28 |
| SOCI | 0.77 | 0.89 | 0.02 | AGRI | 0.45 | 0.54 | 0.00 | CHEM | 0.25 | 0.33 | 0.00 | COMP | 0.13 | 0.14 | 0.20 |

**Note:** Equality of median test determines whether OA and non-OA journals are drawn from populations with the same median.

| | Equality of medians | |
|---|---|---|
| | Me(OA)>=Me(non-OA) | Significant at 1% |
| | Me(OA)>=Me(non-OA) | Significant at 5% |
| | Me(OA)>=Me(non-OA) | Significant at 10% |
| | Me(OA)>=Me(non-OA) | Non-significant |
| | Me(OA)<Me(non-OA) | Non-significant |
| | Me(OA)<Me(non-OA) | Significant at 10% |
| | Me(OA)<Me(non-OA) | Significant at 5% |
| | Me(OA)<Me(non-OA) | Significant at 1% |

**Table 7:** Median values by access type, and equality of median tests for the variable *SJR*, considering those journals firstly included in the Scopus database from 2005 to 2014

| Areas | Journals included in Scopus in 2010-2014 | | | | | Journals included in Scopus in 2005-2009 | | | | |
|---|---|---|---|---|---|---|---|---|---|---|
| | OA | | Non-OA | | Median test | OA | | Non-OA | | Median test |
| | # Journals | Median | # Journals | Median | p-value | # Journals | Median | # Journals | Median | p-value |
| MEDI | 435 | 0.37 | 576 | 0.26 | 0.00 | 311 | 0.46 | 756 | 0.32 | 0.00 |
| PHAR | 47 | 0.42 | 57 | 0.21 | 0.01 | 48 | 0.41 | 116 | 0.33 | 0.49 |
| COMP | 56 | 0.32 | 202 | 0.22 | 0.03 | 59 | 0.31 | 294 | 0.30 | 0.65 |
| ENGI | 104 | 0.24 | 311 | 0.21 | 0.06 | 77 | 0.25 | 412 | 0.26 | 0.91 |
| NEUR | 48 | 0.66 | 43 | 0.41 | 0.17 | 39 | 0.99 | 59 | 0.57 | 0.06 |
| EART | 55 | 0.37 | 71 | 0.28 | 0.21 | 42 | 0.39 | 128 | 0.39 | 0.72 |
| ENVI | 81 | 0.34 | 147 | 0.26 | 0.21 | 55 | 0.41 | 167 | 0.34 | 0.28 |
| CENG | 19 | 0.34 | 60 | 0.30 | 0.39 | 18 | 0.30 | 66 | 0.37 | 0.28 |
| BIOC | 147 | 0.55 | 118 | 0.47 | 0.41 | 104 | 0.75 | 247 | 0.56 | 0.33 |
| VETE | 14 | 0.20 | 12 | 0.18 | 0.43 | 19 | 0.28 | 25 | 0.22 | 0.13 |
| IMMU | 49 | 0.75 | 38 | 0.52 | 0.44 | 23 | 1.05 | 55 | 0.49 | 0.06 |
| DECI | 3 | 0.39 | 15 | 0.20 | 0.52 | 5 | 0.30 | 19 | 0.28 | 0.61 |
| DENT | 17 | 0.32 | 12 | 0.23 | 0.55 | 14 | 0.23 | 13 | 0.25 | 0.84 |
| MATH | 49 | 0.30 | 165 | 0.27 | 0.62 | 43 | 0.37 | 230 | 0.43 | 0.42 |
| HEAL | 28 | 0.24 | 49 | 0.22 | 0.93 | 16 | 0.33 | 44 | 0.23 | 0.08 |
| PHYS | 34 | 0.39 | 68 | 0.38 | 1.00 | 34 | 0.28 | 117 | 0.33 | 0.26 |
| MATE | 40 | 0.31 | 110 | 0.33 | 0.71 | 33 | 0.33 | 145 | 0.26 | 0.03 |
| AGRI | 131 | 0.25 | 137 | 0.30 | 0.71 | 156 | 0.34 | 260 | 0.32 | 0.27 |
| ENER | 21 | 0.30 | 64 | 0.34 | 0.48 | 12 | 0.41 | 57 | 0.32 | 0.49 |
| ARTS | 156 | 0.11 | 661 | 0.11 | 0.32 | 75 | 0.12 | 357 | 0.17 | 0.02 |
| SOCI | 274 | 0.16 | 970 | 0.17 | 0.24 | 195 | 0.20 | 906 | 0.30 | 0.00 |
| CHEM | 27 | 0.34 | 57 | 0.56 | 0.24 | 30 | 0.33 | 69 | 0.29 | 0.94 |
| ECON | 31 | 0.19 | 119 | 0.22 | 0.15 | 28 | 0.19 | 188 | 0.25 | 0.05 |
| NURS | 18 | 0.17 | 49 | 0.26 | 0.11 | 13 | 0.37 | 120 | 0.22 | 0.04 |
| BUSI | 27 | 0.15 | 187 | 0.25 | 0.00 | 25 | 0.19 | 362 | 0.28 | 0.00 |
| PSYC | 39 | 0.19 | 124 | 0.30 | 0.00 | 35 | 0.25 | 157 | 0.36 | 0.02 |
| Total | 1950 | | 4422 | | | 1509 | | 5369 | | |

**Note:** Equality of median test determines whether OA and non-OA journals are drawn from populations with the same median.

| | **Equality of medians** | |
|---|---|---|
| | Me(OA)>=Me(non-OA) | Significant at 1% |
| | Me(OA)>=Me(non-OA) | Significant at 5% |
| | Me(OA)>=Me(non-OA) | Significant at 10% |
| | Me(OA)>=Me(non-OA) | Non-significant |
| | Me(OA)<Me(non-OA) | Non-significant |
| | Me(OA)<Me(non-OA) | Significant at 10% |
| | Me(OA)<Me(non-OA) | Significant at 5% |
| | Me(OA)<Me(non-OA) | Significant at 1% |

**ANNEX 1**: Means, medians (Me), and standard deviations (SD) of the variables *Total Documents*, *References per Document*, and *H Index*, by area and access type

| AREA | Total Documents | | | | | | References per Document | | | | | | H Index | | | | | |
|---|---|---|---|---|---|---|---|---|---|---|---|---|---|---|---|---|---|---|
| | OA | | | Non-OA | | | OA | | | Non-OA | | | OA | | | Non-OA | | |
| | Mean | Me | SD | Mean | Me | SD | Mean | Me | SD | Mean | Me | SD | Mean | Me | SD | Mean | Me | SD |
| AGRI | 143.29 | 44 | 1409.78 | 95.46 | 49 | 177.82 | 36.89 | 34.66 | 17.42 | 36.39 | 35.54 | 23.56 | 20.35 | 12 | 24.99 | 36.00 | 24 | 37.30 |
| ARTS | 33.46 | 18 | 94.50 | 29.27 | 18 | 67.77 | 28.82 | 29.47 | 19.97 | 35.26 | 32.91 | 29.25 | 7.95 | 3 | 24.32 | 12.97 | 5 | 28.46 |
| BIOC | 203.76 | 53 | 1460.40 | 161.29 | 91 | 256.48 | 39.37 | 36.34 | 25.36 | 39.99 | 36.90 | 29.93 | 32.48 | 20 | 36.37 | 64.41 | 48 | 64.63 |
| BUSI | 35.30 | 24 | 35.94 | 47.75 | 29 | 142.84 | 37.08 | 39.61 | 17.72 | 41.88 | 41.99 | 24.38 | 13.63 | 7 | 17.11 | 25.57 | 14 | 29.88 |
| CENG | 223.47 | 63 | 543.51 | 239.46 | 102 | 705.42 | 31.77 | 30.78 | 16.02 | 31.52 | 30.03 | 24.34 | 35.05 | 14 | 50.47 | 45.36 | 28 | 53.12 |
| CHEM | 221.24 | 73.5 | 487.75 | 311.49 | 135 | 674.24 | 33.79 | 31.50 | 17.93 | 36.72 | 32.36 | 31.81 | 30.11 | 16.5 | 41.61 | 57.85 | 37 | 60.05 |
| COMP | 85.02 | 33 | 185.07 | 87.60 | 42 | 136.54 | 31.38 | 30.28 | 16.81 | 28.89 | 27.94 | 19.50 | 20.55 | 12 | 23.28 | 33.25 | 22 | 33.82 |
| DECI | 35.16 | 24.5 | 32.20 | 66.98 | 37 | 95.16 | 33.80 | 31.05 | 15.56 | 33.18 | 30.86 | 18.66 | 18.16 | 11.5 | 19.23 | 34.22 | 21 | 33.54 |
| DENT | 60.88 | 44.5 | 49.70 | 89.98 | 56 | 96.55 | 26.04 | 25.75 | 9.88 | 22.83 | 23.19 | 18.06 | 14.85 | 8 | 16.79 | 36.33 | 29 | 30.43 |
| EART | 64.30 | 27 | 119.12 | 102.55 | 43 | 231.78 | 40.64 | 40.22 | 29.21 | 36.75 | 33.35 | 31.88 | 21.97 | 13 | 25.76 | 33.30 | 19 | 37.35 |
| ECON | 42.88 | 23 | 89.08 | 51.14 | 27 | 172.78 | 35.99 | 34.68 | 14.78 | 35.56 | 34.09 | 23.54 | 9.79 | 6 | 13.05 | 25.53 | 14 | 29.63 |
| ENER | 139.65 | 59.5 | 203.61 | 196.45 | 71 | 377.18 | 30.87 | 27.51 | 15.05 | 24.88 | 23.31 | 21.11 | 20.63 | 11 | 28.01 | 30.69 | 16 | 37.10 |
| ENGI | 139.77 | 44 | 502.09 | 139.57 | 63 | 235.95 | 24.62 | 23.70 | 12.12 | 23.43 | 22.49 | 18.97 | 15.09 | 10 | 17.68 | 30.74 | 17 | 36.33 |
| ENVI | 114.88 | 38 | 213.16 | 111.27 | 43 | 245.37 | 40.79 | 36.92 | 28.10 | 35.96 | 35.35 | 29.03 | 26.16 | 14 | 32.86 | 32.97 | 20 | 36.16 |
| HEAL | 83.68 | 49 | 146.60 | 87.90 | 49 | 117.95 | 27.38 | 28.38 | 12.92 | 27.90 | 28.00 | 15.77 | 18.09 | 11 | 22.57 | 33.71 | 23 | 35.52 |
| IMMU | 163.39 | 60.5 | 369.59 | 137.59 | 83 | 176.46 | 37.94 | 36.62 | 19.31 | 39.69 | 35.52 | 32.01 | 35.18 | 19.5 | 40.93 | 61.07 | 46 | 61.75 |
| MATE | 162.23 | 62 | 393.73 | 224.82 | 96 | 402.78 | 28.66 | 26.29 | 15.28 | 28.74 | 25.85 | 29.93 | 25.88 | 15 | 37.79 | 41.06 | 25 | 46.25 |
| MATH | 94.94 | 37 | 213.22 | 90.15 | 48.5 | 142.94 | 25.24 | 23.58 | 12.59 | 23.42 | 22.17 | 15.00 | 19.20 | 12 | 22.67 | 29.11 | 22 | 27.58 |
| MEDI | 140.01 | 55 | 840.77 | 109.37 | 58 | 173.51 | 27.69 | 26.00 | 20.83 | 24.65 | 22.67 | 22.33 | 24.99 | 14 | 32.48 | 37.53 | 21 | 48.17 |
| MULT | 547.40 | 93 | 1972.77 | 194.97 | 42 | 563.03 | 28.24 | 26.55 | 16.42 | 22.54 | 18.92 | 23.75 | 20.07 | 15 | 19.23 | 44.90 | 13 | 158.87 |
| NEUR | 125.60 | 59.5 | 171.96 | 138.61 | 78 | 187.74 | 41.72 | 42.91 | 23.98 | 40.45 | 38.98 | 25.90 | 31.77 | 22.5 | 38.81 | 63.41 | 51 | 59.38 |
| NURS | 65.97 | 33 | 111.55 | 81.80 | 54 | 93.45 | 25.74 | 27.91 | 14.10 | 23.24 | 22.28 | 18.34 | 17.97 | 11 | 18.84 | 27.79 | 20 | 32.34 |

| | | | | | | | | | | | | | | | | | |
|---|---|---|---|---|---|---|---|---|---|---|---|---|---|---|---|---|---|
| **PHAR** | 108.08 | 47 | 155.24 | 124.24 | 69 | 180.55 | 36.58 | 32.40 | 21.67 | 34.29 | 31.65 | 31.81 | 23.86 | 15 | 25.55 | 42.26 | 28 | 43.79 |
| **PHYS** | 204.40 | 67 | 528.67 | 261.02 | 115 | 456.79 | 35.54 | 27.62 | 44.24 | 32.76 | 27.41 | 36.12 | 30.42 | 17 | 40.67 | 49.68 | 33 | 52.34 |
| **PSYC** | 59.72 | 25 | 147.37 | 57.22 | 37 | 69.84 | 39.63 | 38.88 | 20.37 | 38.27 | 38.40 | 20.40 | 17.28 | 9 | 20.79 | 37.51 | 28 | 36.16 |
| **SOCI** | 36.22 | 22 | 67.72 | 35.93 | 24 | 55.50 | 33.05 | 32.57 | 20.08 | 37.88 | 36.40 | 28.47 | 9.46 | 5 | 14.10 | 17.90 | 10 | 20.96 |
| **VETE** | 78.70 | 49 | 78.96 | 94.19 | 54 | 137.97 | 28.56 | 28.26 | 10.26 | 26.04 | 26.80 | 16.36 | 19.02 | 12.5 | 19.36 | 28.63 | 21 | 25.10 |

**ANNEX 2**: Means, medians (Me), and standard deviations (SD) of the variables *Cites per Document* and *SJR*, by area and access type

| AREA | Cites per Document (3 years) | | | | | | Cites per Document (2 years) | | | | | | SJR | | | | | |
|---|---|---|---|---|---|---|---|---|---|---|---|---|---|---|---|---|---|---|
| | OA | | | Non-OA | | | OA | | | Non-OA | | | OA | | | Non-OA | | |
| | Mean | Me | SD | Mean | Me | SD | Mean | Me | SD | Mean | Me | SD | Mean | Me | SD | Mean | Me | SD |
| AGRI | 1.08 | 0.66 | 1.37 | 1.30 | 0.89 | 1.44 | 1.06 | 0.63 | 1.39 | 1.28 | 0.87 | 1.48 | 0.60 | 0.34 | 0.95 | 0.71 | 0.46 | 0.92 |
| ARTS | 0.27 | 0.09 | 0.58 | 0.43 | 0.15 | 0.92 | 0.27 | 0.08 | 0.66 | 0.42 | 0.13 | 0.95 | 0.22 | 0.11 | 0.38 | 0.33 | 0.13 | 0.66 |
| BIOC | 2.08 | 1.62 | 1.73 | 2.46 | 1.90 | 2.70 | 2.13 | 1.63 | 1.85 | 2.62 | 1.94 | 3.26 | 1.09 | 0.70 | 1.29 | 1.46 | 0.84 | 2.60 |
| BUSI | 0.65 | 0.45 | 0.67 | 1.15 | 0.77 | 1.27 | 0.58 | 0.39 | 0.65 | 1.06 | 0.74 | 1.17 | 0.32 | 0.21 | 0.30 | 0.76 | 0.35 | 1.31 |
| CENG | 1.72 | 0.84 | 2.04 | 1.97 | 1.24 | 2.65 | 1.65 | 0.77 | 1.98 | 2.03 | 1.25 | 3.04 | 0.66 | 0.34 | 0.77 | 0.90 | 0.47 | 1.79 |
| CHEM | 1.45 | 0.78 | 1.79 | 2.19 | 1.50 | 3.34 | 1.38 | 0.73 | 1.71 | 2.20 | 1.51 | 3.26 | 0.53 | 0.28 | 0.65 | 0.93 | 0.53 | 1.68 |
| COMP | 1.20 | 0.81 | 1.22 | 1.45 | 0.98 | 1.52 | 1.13 | 0.72 | 1.22 | 1.41 | 0.98 | 1.46 | 0.57 | 0.32 | 0.82 | 0.74 | 0.45 | 0.85 |
| DECI | 0.99 | 0.70 | 0.99 | 1.38 | 0.95 | 1.34 | 0.88 | 0.58 | 0.84 | 1.32 | 0.90 | 1.26 | 0.63 | 0.34 | 0.62 | 1.07 | 0.66 | 1.31 |
| DENT | 0.83 | 0.57 | 0.88 | 1.09 | 0.93 | 0.99 | 0.82 | 0.57 | 0.87 | 1.20 | 0.97 | 1.32 | 0.37 | 0.26 | 0.32 | 0.54 | 0.46 | 0.42 |
| EART | 1.21 | 0.69 | 1.57 | 1.32 | 0.88 | 1.81 | 1.18 | 0.70 | 1.55 | 1.31 | 0.88 | 1.82 | 0.74 | 0.38 | 1.05 | 0.84 | 0.46 | 1.34 |
| ECON | 0.52 | 0.31 | 0.58 | 1.00 | 0.63 | 1.16 | 0.47 | 0.27 | 0.56 | 0.92 | 0.60 | 1.04 | 0.41 | 0.19 | 0.76 | 0.89 | 0.35 | 1.66 |
| ENER | 1.65 | 0.65 | 3.86 | 1.46 | 0.72 | 2.12 | 1.64 | 0.61 | 4.09 | 1.43 | 0.84 | 1.98 | 0.74 | 0.28 | 1.71 | 0.73 | 0.38 | 0.93 |
| ENGI | 0.89 | 0.61 | 1.01 | 1.20 | 0.70 | 1.74 | 0.88 | 0.56 | 1.29 | 1.21 | 0.70 | 1.92 | 0.40 | 0.26 | 0.53 | 0.64 | 0.34 | 1.07 |
| ENVI | 1.52 | 0.90 | 2.15 | 1.37 | 0.95 | 1.48 | 1.53 | 0.89 | 2.36 | 1.35 | 0.94 | 1.51 | 0.76 | 0.45 | 1.10 | 0.71 | 0.46 | 0.86 |
| HEAL | 1.07 | 0.76 | 1.16 | 1.10 | 0.88 | 0.98 | 1.11 | 0.68 | 1.65 | 1.15 | 0.92 | 1.06 | 0.53 | 0.35 | 0.82 | 0.60 | 0.43 | 0.56 |
| IMMU | 2.14 | 1.61 | 1.78 | 2.46 | 1.90 | 3.14 | 2.16 | 1.52 | 1.91 | 2.70 | 1.91 | 3.93 | 1.20 | 0.74 | 1.37 | 1.54 | 0.92 | 2.97 |
| MATE | 1.45 | 0.78 | 2.01 | 1.71 | 0.96 | 2.81 | 1.51 | 0.76 | 2.34 | 1.73 | 0.97 | 2.97 | 0.63 | 0.34 | 1.00 | 0.82 | 0.41 | 1.61 |
| MATH | 0.86 | 0.57 | 1.04 | 1.02 | 0.70 | 1.17 | 0.81 | 0.53 | 1.02 | 0.99 | 0.70 | 1.07 | 0.64 | 0.36 | 0.90 | 0.89 | 0.60 | 1.02 |
| MEDI | 1.35 | 0.90 | 1.37 | 1.33 | 0.81 | 2.07 | 1.45 | 0.99 | 1.60 | 1.49 | 0.86 | 2.53 | 0.69 | 0.40 | 0.86 | 0.78 | 0.39 | 1.50 |
| MULT | 0.73 | 0.48 | 0.98 | 0.83 | 0.27 | 2.26 | 0.70 | 0.45 | 0.95 | 1.06 | 0.25 | 3.50 | 0.31 | 0.22 | 0.38 | 0.75 | 0.19 | 2.92 |
| NEUR | 2.35 | 2.12 | 1.73 | 2.28 | 1.98 | 1.99 | 2.40 | 2.03 | 1.91 | 2.47 | 2.02 | 2.60 | 1.34 | 0.98 | 1.25 | 1.37 | 1.03 | 1.80 |
| NURS | 1.00 | 0.58 | 1.12 | 0.94 | 0.62 | 1.06 | 1.12 | 0.54 | 1.66 | 1.00 | 0.64 | 1.14 | 0.50 | 0.27 | 0.64 | 0.51 | 0.32 | 0.59 |
| PHAR | 1.58 | 1.09 | 1.60 | 1.72 | 1.27 | 2.07 | 1.57 | 1.08 | 1.65 | 1.84 | 1.35 | 2.35 | 0.60 | 0.39 | 0.61 | 0.75 | 0.47 | 1.01 |
| **PHYS** | 1.50 | 0.73 | 2.38 | 1.74 | 1.02 | 3.01 | 1.57 | 0.74 | 2.74 | 1.83 | 1.05 | 3.29 | 0.79 | 0.33 | 1.39 | 0.98 | 0.51 | 2.10 |
| **PSYC** | 0.87 | 0.49 | 0.89 | 1.44 | 1.07 | 1.61 | 0.87 | 0.48 | 0.91 | 1.38 | 1.03 | 1.57 | 0.47 | 0.23 | 0.48 | 0.85 | 0.54 | 1.04 |
| **SOCI** | 0.47 | 0.22 | 0.69 | 0.74 | 0.45 | 0.91 | 0.46 | 0.19 | 0.75 | 0.71 | 0.43 | 0.89 | 0.31 | 0.17 | 0.42 | 0.49 | 0.26 | 0.67 |
| **VETE** | 0.65 | 0.47 | 0.55 | 0.74 | 0.53 | 0.71 | 0.62 | 0.43 | 0.55 | 0.78 | 0.54 | 0.76 | 0.36 | 0.29 | 0.26 | 0.46 | 0.30 | 0.39 |

**ANNEX 3**: Means, medians (Me), and standard deviations (SD) of the variable *SJR Best Quartile*, by area and access type

| AREA | SJR Best Quartile | | | | | |
|------|------|------|------|------|------|------|
|      | OA | | | Non-OA | | |
|      | *Mean* | *Me* | *SD* | *Mean* | *Me* | *SD* |
| **AGRI** | 2.58 | 3 | 1.03 | 2.37 | 2 | 1.14 |
| **ARTS** | 2.63 | 3 | 1.11 | 2.32 | 2 | 1.13 |
| **BIOC** | 2.55 | 3 | 1.10 | 2.43 | 2 | 1.12 |
| **BUSI** | 2.86 | 3 | 0.99 | 2.32 | 2 | 1.08 |
| **CENG** | 2.46 | 3 | 1.01 | 2.35 | 2 | 1.10 |
| **CHEM** | 2.78 | 3 | 1.01 | 2.34 | 2 | 1.09 |
| **COMP** | 2.55 | 3 | 1.06 | 2.41 | 2 | 1.12 |
| **DECI** | 2.66 | 3 | 1.00 | 2.41 | 2 | 1.11 |
| **DENT** | 2.85 | 3 | 1.08 | 2.32 | 2 | 1.12 |
| **EART** | 2.55 | 3 | 1.04 | 2.39 | 2 | 1.11 |
| **ECON** | 2.91 | 3 | 0.92 | 2.43 | 2 | 1.12 |
| **ENER** | 2.58 | 3 | 0.84 | 2.36 | 2 | 1.11 |
| **ENGI** | 2.54 | 3 | 0.87 | 2.36 | 2 | 1.10 |
| **ENVI** | 2.43 | 2 | 1.05 | 2.41 | 2 | 1.12 |
| **HEAL** | 2.66 | 3 | 1.00 | 2.38 | 2 | 1.14 |
| **IMMU** | 2.53 | 3 | 1.12 | 2.43 | 2 | 1.12 |
| **MATE** | 2.45 | 3 | 0.94 | 2.32 | 2 | 1.08 |
| **MATH** | 2.88 | 3 | 1.01 | 2.39 | 2 | 1.12 |
| **MEDI** | 2.47 | 2 | 1.06 | 2.52 | 3 | 1.15 |
| **MULT** | 2.27 | 2 | 1.01 | 2.58 | 3 | 1.15 |
| **NEUR** | 2.41 | 2 | 1.15 | 2.49 | 2 | 1.10 |
| **NURS** | 2.63 | 3 | 1.19 | 2.39 | 2 | 1.11 |
| **PHAR** | 2.36 | 2 | 1.08 | 2.44 | 2 | 1.11 |
| **PHYS** | 2.83 | 3 | 1.06 | 2.39 | 2 | 1.10 |
| **PSYC** | 2.99 | 3 | 1.00 | 2.41 | 2 | 1.12 |
| **SOCI** | 2.81 | 3 | 1.05 | 2.36 | 2 | 1.12 |
| **VETE** | 2.65 | 2.5 | 0.95 | 2.43 | 2 | 1.16 |